



\font\twelverm=cmr10 scaled 1200    \font\twelvei=cmmi10 scaled 1200
\font\twelvesy=cmsy10 scaled 1200   \font\twelveex=cmex10 scaled 1200
\font\twelvebf=cmbx10 scaled 1200   \font\twelvesl=cmsl10 scaled 1200
\font\twelvett=cmtt10 scaled 1200   \font\twelveit=cmti10 scaled 1200

\skewchar\twelvei='177   \skewchar\twelvesy='60


\def\twelvepoint{\normalbaselineskip=12.4pt
  \abovedisplayskip 12.4pt plus 3pt minus 9pt
  \belowdisplayskip 12.4pt plus 3pt minus 9pt
  \abovedisplayshortskip 0pt plus 3pt
  \belowdisplayshortskip 7.2pt plus 3pt minus 4pt
  \smallskipamount=3.6pt plus1.2pt minus1.2pt
  \medskipamount=7.2pt plus2.4pt minus2.4pt
  \bigskipamount=14.4pt plus4.8pt minus4.8pt
  \def\rm{\fam0\twelverm}          \def\it{\fam\itfam\twelveit}%
  \def\sl{\fam\slfam\twelvesl}     \def\bf{\fam\bffam\twelvebf}%
  \def\mit{\fam 1}                 \def\cal{\fam 2}%
  \def\tt{\twelvett}
  \textfont0=\twelverm   \scriptfont0=\tenrm   \scriptscriptfont0=\sevenrm
  \textfont1=\twelvei    \scriptfont1=\teni    \scriptscriptfont1=\seveni
  \textfont2=\twelvesy   \scriptfont2=\tensy   \scriptscriptfont2=\sevensy
  \textfont3=\twelveex   \scriptfont3=\twelveex  \scriptscriptfont3=\twelveex
  \textfont\itfam=\twelveit
  \textfont\slfam=\twelvesl
  \textfont\bffam=\twelvebf \scriptfont\bffam=\tenbf
  \scriptscriptfont\bffam=\sevenbf
  \normalbaselines\rm}



\def\beginlinemode{\endmode
  \begingroup\parskip=0pt \obeylines\def\\{\par}\def\endmode{\par\endgroup}}
\def\beginparmode{\endmode
  \begingroup \def\endmode{\par\endgroup}}
\let\endmode=\par
{\obeylines\gdef\
{}}
\def\singlespace{\baselineskip=\normalbaselineskip}

\def\oneandahalfspace{\baselineskip=\normalbaselineskip
  \multiply\baselineskip by 3 \divide\baselineskip by 2}
\def\doublespace{\baselineskip=\normalbaselineskip \multiply\baselineskip by 2}

\newcount\firstpageno
\firstpageno=2
\footline={\ifnum\pageno<\firstpageno{\hfil}\else{\hfil\twelverm\folio\hfil}\fi}
\let\rawfootnote=\footnote		
\def\footnote#1#2{{\rm\singlespace\parindent=0pt\rawfootnote{#1}{#2}}}
\def\raggedcenter{\leftskip=4em plus 12em \rightskip=\leftskip
  \parindent=0pt \parfillskip=0pt \spaceskip=.3333em \xspaceskip=.5em
  \pretolerance=9999 \tolerance=9999
  \hyphenpenalty=9999 \exhyphenpenalty=9999 }
\def\dateline{\rightline{\ifcase\month\or
  January\or February\or March\or April\or May\or June\or
  July\or August\or September\or October\or November\or December\fi
  \space\number\year}}
\def\received{\vskip 3pt plus 0.2fill
 \centerline{\sl (Received\space\ifcase\month\or
  January\or February\or March\or April\or May\or June\or
  July\or August\or September\or October\or November\or December\fi
  \qquad, \number\year)}}


\hsize=6.5truein
\vsize=8.9truein
\parskip=\medskipamount
\twelvepoint		
\doublespace		
\overfullrule=0pt	


\def\preprintno#1{
 \rightline{\rm #1}}	

\def\title			
  {\null\vskip 3pt plus 0.2fill
   \beginlinemode \doublespace \raggedcenter \bf}

\def\author			
  {\vskip 4pt plus 0.2fill \beginlinemode
   \singlespace \raggedcenter}

\def\affil			
  {\vskip 3pt plus 0.1fill \beginlinemode
   \oneandahalfspace \raggedcenter \sl}

\def\abstract			
  {\vskip 3pt plus 0.3fill \beginparmode
   \doublespace \narrower ABSTRACT: }

\def\endtitlepage		
  {\endpage			
   \body}

\def\body			
  {\beginparmode}		

\def\head#1{			
  \filbreak\vskip 0.5truein	
  {\immediate\write16{#1}
   \line{\bf{#1}\hfil}\par}
   \nobreak\vskip 0.25truein\nobreak}

\def\refto#1{$^{#1}$}		

\def\references			
  {\head{References}		
   \beginparmode
   \frenchspacing \parindent=0pt \leftskip=0truecm
   \parskip=8pt plus 3pt \everypar{\hangindent=\parindent}}

\gdef\refis#1{\indent\hbox to 0pt{\hss#1.~}}	

\gdef\journal#1, #2, #3, 1#4#5#6{		
    {\sl #1~}{\bf #2}, #3, (1#4#5#6)}		

\def\refstylenp{		
  \gdef\refto##1{ [##1]}				
  \gdef\refis##1{\indent\hbox to 0pt{\hss##1)~}}	
  \gdef\journal##1, ##2, ##3, ##4 {			
     {\sl ##1~}{\bf ##2~}(##3) ##4 }}

\def\refstyleprnp{		
  \gdef\refto##1{ [##1]}				
  \gdef\refis##1{\indent\hbox to 0pt{\hss##1)~}}	
  \gdef\journal##1, ##2, ##3, 1##4##5##6{		
    {\sl ##1~}{\bf ##2~}(1##4##5##6) ##3}}

\def\endreferences{\body}

\def\figurecaptions		
  {
   \beginparmode
   \head{Figure Captions}
}

\def\endpage			
  {\vfill\eject}

\def\endpaper			
  {\endmode\vfill\supereject}

\def\endit
  {\endpaper\end}


\def\ref#1{Ref.\thinspace#1}			

\def\Ref#1{Ref.\thinspace#1}			
\def\fig#1{Fig.\thinspace#1}
\def\figs#1{Figs.\thinspace#1}

\def\frac#1#2{{#1 \over #2}}
\def\ffrac#1#2{\textstyle{#1\over#2}\displaystyle}  
\def\eg{{\it e.g.,\ }}

\def\sla{\raise.15ex\hbox{$/$}\kern-.57em}
\def\leaderfill{\leaders\hbox to 1em{\hss.\hss}\hfill}
\def\twiddle{\lower.9ex\rlap{$\kern-.1em\scriptstyle\sim$}}
\def\bigtwiddle{\lower1.ex\rlap{$\sim$}}
\def\gtwid{\mathrel{\raise.3ex\hbox{$>$\kern-.75em\lower1ex\hbox{$\sim$}}}}
\def\ltwid{\mathrel{\raise.3ex\hbox{$<$\kern-.75em\lower1ex\hbox{$\sim$}}}}
\def\square{\kern1pt\vbox{\hrule height 1.2pt\hbox{\vrule width 1.2pt\hskip 3pt
   \vbox{\vskip 6pt}\hskip 3pt\vrule width 0.6pt}\hrule height 0.6pt}\kern1pt}
\def\ucsb{Department of Physics\\University of California\\
Santa Barbara CA 93106}


\def\(#1){Eq.\thinspace(\call{#1})}
\def\call#1{{#1}}

\abovedisplayskip=18pt plus 3pt minus 9pt
\belowdisplayskip=18pt plus 3pt minus 9pt



\def\rhs{right hand side}

\mathchardef\duall="0414


\def\ev#1{\langle #1\rangle}

\def\Ln{\mathop{\rm Ln}\nolimits}

\def\Z#1{{\rm Z}_#1}  

\def\Z2{${\rm Z}_2$}

\def\one{{\bf 1}}      

\def\ope{operator product expansion}

\def\rg{renormalization group}

\def\cf{correlation function}

\def\cfs{\cf s}

\def\cft{conformal field theory}

\def\cfts{conformal field theories}

\def\eg{{\it e.g.}}

\def\uhp{upper half plane}

\def\a{\alpha}
\def\b{\beta}
\def\g{\gamma}
\def\d{\delta}

\def\G{\Gamma}

\refstylenp

\catcode`@=11
\newcount\r@fcount \r@fcount=0
\newcount\r@fcurr
\immediate\newwrite\reffile
\newif\ifr@ffile\r@ffilefalse
\def\w@rnwrite#1{\ifr@ffile\immediate\write\reffile{#1}\fi\message{#1}}

\def\writer@f#1>>{}
\def\referencefile{
  \r@ffiletrue\immediate\openout\reffile=\jobname.ref%
  \def\writer@f##1>>{\ifr@ffile\immediate\write\reffile%
    {\noexpand\refis{##1} = \csname r@fnum##1\endcsname = %
     \expandafter\expandafter\expandafter\strip@t\expandafter%
     \meaning\csname r@ftext\csname r@fnum##1\endcsname\endcsname}\fi}%
  \def\strip@t##1>>{}}

\def\citeall#1{\xdef#1##1{#1{\noexpand\cite{##1}}}}
\def\cite#1{\each@rg\citer@nge{#1}}	

\def\each@rg#1#2{{\let\thecsname=#1\expandafter\first@rg#2,\end,}}
\def\first@rg#1,{\thecsname{#1}\apply@rg}	
\def\apply@rg#1,{\ifx\end#1\let\next=\relax
\else,\thecsname{#1}\let\next=\apply@rg\fi\next}

\def\citer@nge#1{\citedor@nge#1-\end-}	
\def\citer@ngeat#1\end-{#1}
\def\citedor@nge#1-#2-{\ifx\end#2\r@featspace#1 
  \else\citel@@p{#1}{#2}\citer@ngeat\fi}	
\def\citel@@p#1#2{\ifnum#1>#2{\errmessage{Reference range #1-#2\space is bad.}%
    \errhelp{If you cite a series of references by the notation M-N, then M and
    N must be integers, and N must be greater than or equal to M.}}\else%
 {\count0=#1\count1=#2\advance\count1
by1\relax\expandafter\r@fcite\the\count0,%
  \loop\advance\count0 by1\relax
    \ifnum\count0<\count1,\expandafter\r@fcite\the\count0,%
  \repeat}\fi}

\def\r@featspace#1#2 {\r@fcite#1#2,}	
\def\r@fcite#1,{\ifuncit@d{#1}
    \newr@f{#1}%
    \expandafter\gdef\csname r@ftext\number\r@fcount\endcsname%
                     {\message{Reference #1 to be supplied.}%
                      \writer@f#1>>#1 to be supplied.\par}%
 \fi%
 \csname r@fnum#1\endcsname}
\def\ifuncit@d#1{\expandafter\ifx\csname r@fnum#1\endcsname\relax}%
\def\newr@f#1{\global\advance\r@fcount by1%
    \expandafter\xdef\csname r@fnum#1\endcsname{\number\r@fcount}}

\let\r@fis=\refis			
\def\refis#1#2#3\par{\ifuncit@d{#1}
   \newr@f{#1}%
   \w@rnwrite{Reference #1=\number\r@fcount\space is not cited up to now.}\fi%
  \expandafter\gdef\csname r@ftext\csname r@fnum#1\endcsname\endcsname%
  {\writer@f#1>>#2#3\par}}

\def\ignoreuncited{
   \def\refis##1##2##3\par{\ifuncit@d{##1}%
     \else\expandafter\gdef\csname r@ftext\csname
r@fnum##1\endcsname\endcsname%
     {\writer@f##1>>##2##3\par}\fi}}

\def\r@ferr{\endreferences\errmessage{I was expecting to see
\noexpand\endreferences before now;  I have inserted it here.}}
\let\r@ferences=\references
\def\references{\r@ferences\def\endmode{\r@ferr\par\endgroup}}

\let\endr@ferences=\endreferences
\def\endreferences{\r@fcurr=0
  {\loop\ifnum\r@fcurr<\r@fcount
    \advance\r@fcurr by 1\relax\expandafter\r@fis\expandafter{\number\r@fcurr}%
    \csname r@ftext\number\r@fcurr\endcsname%
  \repeat}\gdef\r@ferr{}\endr@ferences}


\let\r@fend=\endpaper\gdef\endpaper{\ifr@ffile
\immediate\write16{Cross References written on []\jobname.REF.}\fi\r@fend}

\catcode`@=12

\citeall\refto		
\citeall\ref		%
\citeall\Ref		%

\catcode`@=11
\newcount\tagnumber\tagnumber=0

\immediate\newwrite\eqnfile
\newif\if@qnfile\@qnfilefalse
\def\write@qn#1{}
\def\writenew@qn#1{}
\def\w@rnwrite#1{\write@qn{#1}\message{#1}}
\def\@rrwrite#1{\write@qn{#1}\errmessage{#1}}

\def\taghead#1{\gdef\t@ghead{#1}\global\tagnumber=0}
\def\t@ghead{}

\expandafter\def\csname @qnnum-3\endcsname
  {{\t@ghead\advance\tagnumber by -3\relax\number\tagnumber}}
\expandafter\def\csname @qnnum-2\endcsname
  {{\t@ghead\advance\tagnumber by -2\relax\number\tagnumber}}
\expandafter\def\csname @qnnum-1\endcsname
  {{\t@ghead\advance\tagnumber by -1\relax\number\tagnumber}}
\expandafter\def\csname @qnnum0\endcsname
  {\t@ghead\number\tagnumber}
\expandafter\def\csname @qnnum+1\endcsname
  {{\t@ghead\advance\tagnumber by 1\relax\number\tagnumber}}
\expandafter\def\csname @qnnum+2\endcsname
  {{\t@ghead\advance\tagnumber by 2\relax\number\tagnumber}}
\expandafter\def\csname @qnnum+3\endcsname
  {{\t@ghead\advance\tagnumber by 3\relax\number\tagnumber}}

\def\equationfile{%
  \@qnfiletrue\immediate\openout\eqnfile=\jobname.eqn%
  \def\write@qn##1{\if@qnfile\immediate\write\eqnfile{##1}\fi}
  \def\writenew@qn##1{\if@qnfile\immediate\write\eqnfile
    {\noexpand\tag{##1} = (\t@ghead\number\tagnumber)}\fi}
}

\def\callall#1{\xdef#1##1{#1{\noexpand\call{##1}}}}
\def\call#1{\each@rg\callr@nge{#1}}

\def\each@rg#1#2{{\let\thecsname=#1\expandafter\first@rg#2,\end,}}
\def\first@rg#1,{\thecsname{#1}\apply@rg}
\def\apply@rg#1,{\ifx\end#1\let\next=\relax%
\else,\thecsname{#1}\let\next=\apply@rg\fi\next}

\def\callr@nge#1{\calldor@nge#1-\end-}
\def\callr@ngeat#1\end-{#1}
\def\calldor@nge#1-#2-{\ifx\end#2\@qneatspace#1 %
  \else\calll@@p{#1}{#2}\callr@ngeat\fi}
\def\calll@@p#1#2{\ifnum#1>#2{\@rrwrite{Equation range #1-#2\space is bad.}
\errhelp{If you call a series of equations by the notation M-N, then M and
N must be integers, and N must be greater than or equal to M.}}\else%
 {\count0=#1\count1=#2\advance\count1
by1\relax\expandafter\@qncall\the\count0,%
  \loop\advance\count0 by1\relax%
    \ifnum\count0<\count1,\expandafter\@qncall\the\count0,%
  \repeat}\fi}

\def\@qneatspace#1#2 {\@qncall#1#2,}
\def\@qncall#1,{\ifunc@lled{#1}{\def\next{#1}\ifx\next\empty\else
  \w@rnwrite{Equation number \noexpand\(>>#1<<) has not been defined yet.}
  >>#1<<\fi}\else\csname @qnnum#1\endcsname\fi}

\let\eqnono=\eqno
\def\eqno(#1){\tag#1}
\def\tag#1$${\eqnono(\displayt@g#1 )$$}

\def\aligntag#1$${\gdef\tag##1\\{&(\displayt@g##1 )\cr}\eqalignno{#1\\}$$
  \gdef\tag##1$${\eqnono(\displayt@g##1 )$$}}

\def\eqalignno#1{\displ@y \tabskip\centering
  \halign to\displaywidth{\hfil$\displaystyle{##}$\tabskip\z@skip
    &$\displaystyle{{}##}$\hfil\tabskip\centering
    &\llap{$\displayt@gpar##$}\tabskip\z@skip\crcr
    #1\crcr}}

\def\displayt@gpar(#1){(\displayt@g#1 )}

\def\displayt@g#1 {\rm\ifunc@lled{#1}\global\advance\tagnumber by1
        {\def\next{#1}\ifx\next\empty\else\expandafter
        \xdef\csname @qnnum#1\endcsname{\t@ghead\number\tagnumber}\fi}%
  \writenew@qn{#1}\t@ghead\number\tagnumber\else
        {\edef\next{\t@ghead\number\tagnumber}%
        \expandafter\ifx\csname @qnnum#1\endcsname\next\else
        \w@rnwrite{Equation \noexpand\tag{#1} is a duplicate number.}\fi}%
  \csname @qnnum#1\endcsname\fi}

\def\ifunc@lled#1{\expandafter\ifx\csname @qnnum#1\endcsname\relax}

\let\@qnend=\end\gdef\end{\if@qnfile
\immediate\write16{Equation numbers written on []\jobname.EQN.}\fi\@qnend}

\catcode`@=12


\preprintno{UCSBTH--91--56}
\title
Critical Percolation in Finite Geometries
\author
John L. Cardy
\affil\ucsb
\abstract
The methods of conformal field theory are used to compute the
crossing probabilities between segments of the boundary of a
compact two-dimensional region at the percolation threshold.
These probabilities are shown to be invariant not only under
changes of scale, but also under mappings of the region which
are conformal in the interior and continuous on the boundary.
This is a larger invariance than that expected for generic
critical systems. Specific predictions are presented for
the crossing probability between opposite sides of a rectangle, and are
compared with recent numerical work. The agreement is excellent.
\vfill
\line{PACS Nos: 64.60.Ak, 05.50.+k\hfil}
\eject

Conformal field theory
has been very successful in determining universal quantities associated
with two-dimensional isotropic systems at their critical
points\refto{CARDYDL,CARDYFSS}.
The range of predictions which can be made appears to be bounded by the
enthusiasm and industriousness of the theorist rather than by any intrinsic
limitations of the theory. However, the underlying assumptions of
conformal field theory, and their appropriateness for describing
the scaling limit of critical lattice systems, are not rigorously
founded, and it remains important to perform precise numerical tests
of the theory whenever possible.

Recently\refto{LANG}, extensive numerical work has been carried out
to estimate crossing probabilities in rectangular geometries for critical
percolation in very large but finite lattices, with the principal
aim of establishing their universality between different models.
Percolation provides an important test of the ideas of conformal field
theory
because large-scale
numerical simulations are more readily performed. In this
letter we consider the general problem of crossing probabilities in the
language of conformal field theory, and derive exact expressions which
may be compared with the numerical work.

The most familiar way to think about percolation as a critical
phenomenon is through the $q\to1$ limit of the $q$-state
Potts model\refto{KF}.  In that model, spins $s(r)$ at the
sites of the lattice are allowed to be in one of $q$
possible states $(\a,\b,\ldots)$, and the
partition function is the trace of a product over links of the form
$$
Z=\prod_{(r,r')}\left(1+x\d_{s(r),s(r')}\right)
\eqno()
$$
The terms in the expansion in powers of $Z$ in powers of $x$
are in 1-1 correspondence with configurations of bonds appearing in the
bond percolation problem, and in the limit $q\to1$ they are weighted
appropriately if $x=p/(1-p)$.
Two sites in the same cluster are necessarily in the
same state of the Potts model.
Consider now two disjoint segments $S_1$ and $S_2$
of the piecewise differentiable
boundary of a simply connected compact region. Let $Z_{\a\b}$
be the partition function of the $q$-state Potts model with the
constraint that all spins at lattice sites on $S_1$ are fixed in the state
$\a$, and all the spins on $S_2$ are fixed in the state $\b$. The rest
of the boundary spins are unrestricted. Then the crossing probability
between $S_1$ and $S_2$ is given by
$$
\pi(S_1,S_2)=\lim_{q\to1}\left(Z_{\a\a}-Z_{\a\b}\right)
\eqno()
$$
where, in the second term $\a\not=\b$. In fact, in the limit when
$q=1$, the first term is unity.

The interior of the compact region may be mapped conformally to the
\uhp, so that the boundary is mapped onto the real axis. If there are
corners on the boundary, the map will be singular but continuous at
these points. Conformal field theory relates the partition functions
in the two geometries, in a manner to be described later. Thus, if
the images of $S_1$ and $S_2$ are the intervals $(x_1,x_2)$ and
$(x_3,x_4)$ respectively (where we may assume that the $x_i$ are
placed in increasing order), the problem reduces to that of finding
the respective partition functions $Z_{\a\a}$ and $Z_{\a\b}$ in this
geometry.

The study of boundary conditions in conformal field
theory\refto{CARDYSURF,CARDYBC} shows that,
for a particular theory, there is a given set of boundary conditions
consistent with the conformal symmetry of the theory. In general they
correspond to the possible fixed points of the \rg\ in the semi-infinite
system: thus a generic boundary condition becomes equivalent in the
continuum limit to one of
those allowed by conformal symmetry.
In addition, points on the boundary at which the boundary condition
changes may be identified\refto{CARDYBC}
with points of insertion of boundary operators, that
is, scaling operators of the theory corresponding to highest weight
states of the Virasoro algebra\refto{BPZ,CARDYSURF}.
Situations where more than one change
of boundary condition occurs then correspond to correlation functions
of these boundary operators.
In the case in question, let us denote the boundary condition where
the spins are free by $(f)$, and those where they are fixed in a given
state by $(\a)$. Denote the boundary operator corresponding to a switch
from boundary condition $(i)$ to $(j)$ at the point $x$ by
$\phi_{(i|j)}(x)$. Then the partition functions we need are given
in terms of correlators by
$$
\eqalign{
Z_{\a\a}&=Z_f\,\ev{\phi_{(f|\a)}(x_1)\,\phi_{(\a|f)}(x_2)\,\phi_{(f|\a)}(x_3)
                               \,\phi_{(\a|f)}(x_4)}\cr
Z_{\a\b}&=Z_f\,\ev{\phi_{(f|\a)}(x_1)\,\phi_{(\a|f)}(x_2)\,\phi_{(f|\b)}(x_3)
                               \,\phi_{(\b|f)}(x_4)}\cr}
\eqno(ZZ)
$$
where $Z_f$ is partition function with free boundary conditions all
along the real axis. Note that, in the \uhp, all three partition
functions in general diverge in the infinite volume limit, and \(ZZ)
strictly should be interpreted as being valid only for a large
but finite lattice. However, when $q=1$, $Z_f=1$ identically, and this
problem does not arise.

In order to compute the above \cfs\ using the methods of \cft, we
need to understand to which representations of the Virasoro algebra
the boundary operators belong.
It has been known for some time\refto{DOTFAT,SACLAY} that the critical
$q$-state
Potts model corresponds, in the continuum limit, to a \cft\ with
conformal anomaly number\refto{BPZ} $c=1-6/m(m+1)$, where
$q=4\cos^2(\pi/m+1)$, with $m\geq1$. Thus percolation has $c=0$. This
is consistent with the fact that $c$ is related to the finite-size
corrections to the free energy\refto{CARDYPESCH} in certain geometries,
and the free energy vanishes identically when $q=1$.
However, the problem of boundary operator assignment has not been addressed
so far, except for the cases
$q=2$ and $q=3$\refto{CARDYOPC,CARDYBC,SB}. However, it is
not difficult to determine the assignment for the operators $\phi_{(f|\a)}$.
For minimal \cfts, all the scaling operators have the property that
their corresponding representations contain null states\refto{BPZ}.
This has the
consequence that the allowed values of their scaling dimensions are
given by the Kac formula
$$
h=h_{r,s}={\big(r(m+1)-sm\big)^2-1\over 4m(m+1)}
\eqno(KAC)
$$
where $r$ and $s$ are positive integer are positive integers.
In addition, the correlators involving these operators
obey differential equations of order at most $rs$. For unitary models,
for example those with positive definite Boltzmann weights, all
allowed operators must be of this type. Although this condition is not
applicable to the $q$-state Potts model for general $q$, the fact that
it does apply for $q=2$ and $q=3$ suggests that those
operators whose position $(r,s)$ in the Kac table does not appear to
change as a function of $c(q)$ do correspond to representations with
null states even in the non-unitary case.
Indeed, it was conjectured in \ref{CARDYSURF} that the spin operator of the
Potts model, when inserted at a boundary with free boundary conditions,
corresponds to $(r,s)=(1,3)$. This agrees with known results for
$q=2,3$\refto{CARDYOPC,CARDYBC,SB},
and is also consistent with the known assignment of operators in
the bulk. It gives a prediction for the case $q=1$ which agrees with
numerical estimates to within their accuracy\refto{DEBELL}.
There are $(q-1)$ independent
such spin operators.

The continuum limit of duality symmetry\refto{POTTS}
for the critical $q$-state Potts model
maps the free boundary condition $(f)$ onto a fixed boundary condition
$(\a)$. Exactly which state $\a$ is chosen is arbitrary, since just
one spin on the boundary has to be assigned a given value in order to
make the duality mapping 1-1.
An insertion of the spin operator at the point $x$
on the free boundary is mapped
into an insertion of the disorder operator $\phi_{(\a|\b)}(x)$
where $\b\not=\a$. There are just $(q-1)$ such operators, for a fixed
$\a$. This duality symmetry implies that that the correlators of
$\phi{(\a|\b)}$ are simply related to those of the boundary spin
operator with free boundary conditions, and hence that it also
corresponds to $(r,s)=(1,3)$ in the Kac table. However, we are
interested in the operators $\phi_{(\a|f)}$. Consider the
insertion of two such operators $\phi_{(\a|f)}(x)\phi_{(f|\b)}(x')$
as the points $x$, $x'$ approach each other. This is given by
the \ope, which symbolically must have the structure
$$
\phi_{(\a|f)}\cdot\phi_{(f|\b)}\sim\d_{\a\b}\,\one+\phi_{(\a|\b)}+\cdots
\eqno(OPE)
$$
where $\one$ is the identity operator (no change in boundary condition).
According to the fusion rules of \cft\refto{BPZ},
there is one such operator in the
Kac table which has such a simple \ope\ with itself, namely
$(r,s)=(1,2)$. We therefore conjecture that this is the correct
assignment for the operators $\phi_{(f|\a)}$, for general $q$. This
agrees with the known results for $q=2$ and $q=3$\refto{CARDYOPC,CARDYBC,SB}.
It implies that
the correlators involving these operators satisfy second order
differential equations. From the Kac formula \(KAC), we see that, in the
limit $q\to1$, their scaling dimensions are given by
$$
h=h_{1,2}(0)=0\quad.
\eqno()
$$
This vanishing of the scaling dimension will turn out to have remarkable
consequences when the result is transformed back into the original
geometry. In fact, it has a natural explanation. Consider a compact region on
whose boundary there is a single segment $S_1$ on which the Potts spins
are fixed into the state $\a$. On the remainder of the boundary, the
spins are free. In the limit $q\to1$, the partition function in this
geometry is equal to unity, and equal to $Z_f$, since in either case any
spin can only be in a single state. But the ratio of these partition
functions is equal to the correlation function
$\ev{\phi_{(f|\a)}\phi_{(\a|f)}}$, which in general will scale like
distance to the power $2h$. This is only consistent if $h=0$.
However, the form of the four-point functions, although simplified by
this result, is nontrivial.
Consider the half plane geometry, when
the points lie along the real axis. Conformal invariance
implies\refto{BPZ} that they are of the form $F(\eta)$,
depending only on the invariant
cross-ratio $\eta=(x_4-x_3)(x_2-x_1)/(x_3-x_1)(x_4-x_2)$. The absence
of other prefactors multiplying $F$ is a consequence of $h=0$. The
fact that the correlators satisfy second order differential equations
implies that $F(\eta)$ satisfies a Riemann equation, whose general
solution is\refto{BPZ,CARDYSURF}
$$
F=P\left\{\matrix{0&\infty&1&{}\cr
                0&-4h_{1,2}&0&\eta\cr
                h_{1,3}&-4h_{1,2}+h_{1,3}&h_{1,3}&{}\cr}\right\}
\eqno()
$$
Which solution is chosen depends on whether we calculate $Z_{\a\a}$
or $Z_{\a\b}$. Although is straightforward to solve this problem for
arbitrary $q$, we restrict ourselves to $q=1$ for simplicity.
In that limit, one of the solutions of the Riemann equation reduces
to a constant, and the second solution is
proportional to $\eta^{1/3}{}_2F_1(\ffrac13,\ffrac23,\ffrac43;\eta)$.
The combination corresponding to $Z_{\a\b}$ is determined by the
requirement that as $(x_3-x_2)\to0$, that is $\eta\to1$, the
\ope\ \(OPE) requires that the solution vanish like $(1-\eta)^{1/3}$.
In addition, in the opposite limit $\eta\to0$, we expect that
$Z_{\a\b}\to Z_{\a\a}=1$. Using simple identities on hypergeometric
functions, we then find for the crossing probability
$$
\eqalign{
\pi\big((x_1,x_2),(x_3,x_4)\big)&=
{3\G(\ffrac23)\over\G(\ffrac13)^2}
\eta^{1/3}{}_2F_1(\ffrac13,\ffrac23,\ffrac43;\eta)\cr
&=1-
{3\G(\ffrac23)\over\G(\ffrac13)^2}
(1-\eta)^{1/3}{}_2F_1(\ffrac13,\ffrac23,\ffrac43;1-\eta)\cr}\quad.
\eqno(PI)
$$

Now consider the transformation of the \uhp\ onto the interior of
a simply connected compact region by a conformal mapping
$z\to w$. If the
boundary of the region is a differentiable curve, this mapping may
be taken to be
conformal also on the boundary. In that case, correlation functions
of operators on the boundary transform in the standard manner
summarized by the formula
$$
\ev{\phi_1(w_1)\phi_2(w_2)\ldots}=\prod_i|w'(z_i)|^{-h_i}
\ev{\phi_1(z_1)\phi_2(z_2)\ldots}
\eqno(TRANS)
$$
where the correlation functions on the left and \rhs s refer to the
new and the old geometry respectively, and
$h_i$ is the scaling dimension of $\phi_i$. In our case, however,
since $h=0$, no such prefactors arise, and the correlation function
is truly invariant. For a general critical system, the partition function
for a compact region
(without any operator insertions) is not itself scale invariant, but picks
up a factor $(L/L_0)^{ac}$
where $L$ has the dimensions of length and gives the overall size of the
region, $L_0$ is some non-universal microscopic scale (\eg\ the lattice
spacing), and $a$ is geometry dependent\refto{CARDYPESCH}.
However, for the case of percolation, $c=0$ and $Z=1$,
so such effects are absent.
In general, there is a further complication
when the boundary of the compact region is only piecewise
differentiable, and boundary operators happen to sit at the corners.
In this case \(TRANS) does not apply. Instead there appear additional
non-scale invariant factors of the form $(L/L_0)^{-(\pi/\g)h}$, where
$\g$ is the interior angle at the corner.
Such factors have been treated explicitly for the Ising model with
various boundary conditions\refto{KLEBAN}.
However, once again, since $h=0$ for the problem at hand, such factors
are absent. We conclude that crossing probabilities are indeed
invariant under mappings which are conformal in the interior and
are piecewise conformal on the boundary, but that this is not generic
for all critical systems, for example when $q\not=1$.

As an example, consider the case treated in \ref{LANG} of the crossing
probability between opposite sides of a rectangle of aspect ratio $r$.
This is the image of the \uhp\ under a Schwartz-Christoffel
transformation. Taking the points $x_j$ to be at
$(-k^{-1},-1,1,k^{-1})$, the aspect ratio of the rectangle is given by
$r=K(1-k^2)/2K(k^2)$, where $K(u)$ is the complete elliptic integral
of the first kind. The prediction is then that the crossing probability
is given by \(PI), with $\eta=((1-k)/(1+k))^2$. The results of this are
illustrated in \figs{1,2}, and compared with the numerical data obtained in
\ref{LANG} for bond percolation on square lattices with approximately
$4\cdot10^4$ sites. It is seen from \fig{2} that the deviations between
the numerical experiment and the theory are consistent with the internal
scatter of the data, although there appears also to be a systematic
difference which may be due to finite-size effects.

It is possible to generalize the above methods to treat the case of
correlations between different crossing events. As long as the segments
involved are not adjacent, such probabilities may always be related
to correlation functions
of $\phi_{(f|\a)}$ operators, and they should have the
same invariance properties as the simple
crossing probabilities considered here. The fact that they
enjoy these properties, which are not expected to hold for analogous
quantities in generic critical systems, suggests that some of the ideas
of conformal invariance might usefully be reformulated for the
percolation problem without invoking the mapping to the Potts model.

The author is grateful to R. P. Langlands for providing a copy of
\ref{LANG} before publication, and the numerical data shown in \figs{1,2}.
He also thanks T. Spencer and M. Aizenman for stimulating his interest
in this problem, and P. Kleban for communicating the results of
\ref{KLEBAN}. This work was supported by NSF Grant PHY 86-14185.

\head{Figure Captions}

\item{1)} Theory {\it vs.} the numerical data of \ref{LANG} for
the horizontal crossing probabilities $\pi_h(r)$ for rectangles
of aspect ratio $r$. In the figure, $\Ln\big((1-\pi_h)/\pi_h\big)$
is plotted against $\Ln r$. The numerical data is represented by
by points, and the solid curve is the theoretical prediction.

\item{2)} Deviation between numerical estimates and theoretical
predictions of crossing
probabilities $\pi_h(r)$ and $1-\pi_v(r)$.

\references

\refis{CARDYDL} Cardy J L in {\sl Phase Transitions and Critical
Phenomena}, v.11, C. Domb and J. L. Lebowitz eds. (Academic, 1987)

\refis{CARDYFSS} See articles in {\sl Finite-Size Scaling,
Current Physics -- Sources and Comments}, v.2, J. L. Cardy,
ed. (North-Holland, 1988)

\refis{LANG} Langlands R P, Pichet C, Pouliot Ph and Saint-Aubin Y,
{\sl On the Universality of Crossing Probabilities in Two-Dimensional
Percolation}, preprint CRM-1785, October 1991.

\refis{KF} Kasteleyn P W and Fortuin C M 1989 {\sl J. Phys. Soc.
Japan} {\bf 26} {\sl (Supp.)} 11

\refis{CARDYSURF} Cardy J L 1984 {\sl Nucl. Phys.} {\bf B240} 514

\refis{CARDYBC} Cardy J L 1989 {\sl Nucl. Phys.} {\bf B324} 581

\refis{BPZ} Belavin A A, Polyakov A M and Zamolodchikov A B 1984
{\sl Nucl. Phys.} {\bf B241} 333

\refis{DOTFAT} Dotsenko Vl S and Fateev V A 1984 {\sl Nucl. Phys.}
{\bf B240} 312

\refis{SACLAY} Di Francesco P, Saleur H and Zuber J-B 1987 {\sl J.
Stat. Phys.} {\bf 49} 57

\refis{CARDYPESCH} Cardy J L and Peschel I 1988 {\sl Nucl. Phys.}
{\bf B300} 377

\refis{CARDYOPC} Cardy J L 1986 {\sl Nucl. Phys.} {\bf B275} 200

\refis{SB} Saleur H and Bauer M 1989 {\sl Nucl. Phys.} {\bf B320}
591

\refis{DEBELL} De'Bell K and Essam J W 1980 {\sl J. Phys. C}
{\bf 13} 4811

\refis{POTTS} Potts R B 1952 {\sl Proc. Camb. Phil.} {\bf 48} 106

\refis{KLEBAN} Kleban P and Vassileva I, in preparation.

\endreferences

\endit


MathPictureStart
0.02397 0.46679 0.01494 0.07644 [
[(Fig. 1)] 0.5 0.62428 0 -1 Msboxa
[(0.5)] 0.25737 0.00244 0 1 Msboxa
[(1)] 0.49077 0.00244 0 1 Msboxa
[(1.5)] 0.72416 0.00244 0 1 Msboxa
[(2)] 0.95756 0.00244 0 1 Msboxa
[(Ln[r])] 1.00625 0.01494 -1 0 Msboxa
[(2)] 0.01147 0.16781 1 0 Msboxa
[(4)] 0.01147 0.32069 1 0 Msboxa
[(6)] 0.01147 0.47357 1 0 Msboxa
[(Ln[\(1-pi\)/pi])] 0.02397 0.62428 0 -1 Msboxa
[ -0.001 -0.001 0 0 ]
[ 1.001 0.61903 0 0 ]
] MathScale
1 setlinecap
1 setlinejoin
newpath
[ ] 0 setdash
0 setgray
0 setgray
[(Fig. 1)] 0.5 0.62428 0 -1 Mshowa
gsave
gsave
0.002 setlinewidth
0 0.01494 moveto
1 0.01494 lineto
stroke
0.25737 0.00869 moveto
0.25737 0.02119 lineto
stroke
0 setgray
[(0.5)] 0.25737 0.00244 0 1 Mshowa
0.49077 0.00869 moveto
0.49077 0.02119 lineto
stroke
0 setgray
[(1)] 0.49077 0.00244 0 1 Mshowa
0.72416 0.00869 moveto
0.72416 0.02119 lineto
stroke
0 setgray
[(1.5)] 0.72416 0.00244 0 1 Mshowa
0.95756 0.00869 moveto
0.95756 0.02119 lineto
stroke
0 setgray
[(2)] 0.95756 0.00244 0 1 Mshowa
0 setgray
[(Ln[r])] 1.00625 0.01494 -1 0 Mshowa
0.02397 0 moveto
0.02397 0.61803 lineto
stroke
0.01772 0.16781 moveto
0.03022 0.16781 lineto
stroke
0 setgray
[(2)] 0.01147 0.16781 1 0 Mshowa
0.01772 0.32069 moveto
0.03022 0.32069 lineto
stroke
0 setgray
[(4)] 0.01147 0.32069 1 0 Mshowa
0.01772 0.47357 moveto
0.03022 0.47357 lineto
stroke
0 setgray
[(6)] 0.01147 0.47357 1 0 Mshowa
0 setgray
[(Ln[\(1-pi\)/pi])] 0.02397 0.62428 0 -1 Mshowa
grestore
grestore
0 0 moveto
1 0 lineto
1 0.618034 lineto
0 0.618034 lineto
closepath
clip
newpath
0 setgray
gsave
gsave
gsave
gsave
0.004 setlinewidth
0.02381 0.01488 moveto
0.2052 0.07805 lineto
0.30538 0.116 lineto
0.37884 0.1464 lineto
0.43839 0.17321 lineto
0.48921 0.19799 lineto
0.53396 0.22152 lineto
0.57417 0.24423 lineto
0.61082 0.2664 lineto
0.64458 0.2882 lineto
0.67593 0.30972 lineto
0.70523 0.33106 lineto
0.73274 0.35227 lineto
0.75869 0.37338 lineto
0.78325 0.39441 lineto
0.80657 0.4154 lineto
0.82877 0.43634 lineto
0.84996 0.45726 lineto
0.87022 0.47816 lineto
0.88964 0.49904 lineto
0.90828 0.51991 lineto
0.92621 0.54077 lineto
0.94347 0.56162 lineto
0.96012 0.58247 lineto
0.97619 0.60332 lineto
stroke
grestore
grestore
grestore
gsave
gsave
0.008 setlinewidth
0.02397 0.01472 Mdot
0.04719 0.02259 Mdot
0.07058 0.0305 Mdot
0.09366 0.03914 Mdot
0.11718 0.04658 Mdot
0.14284 0.05538 Mdot
0.16371 0.06251 Mdot
0.18667 0.07115 Mdot
0.20949 0.07906 Mdot
0.23215 0.0877 Mdot
0.25688 0.09666 Mdot
0.27983 0.10527 Mdot
0.30453 0.11519 Mdot
0.32604 0.12334 Mdot
0.35079 0.13348 Mdot
0.37561 0.14394 Mdot
0.39708 0.1531 Mdot
0.42002 0.16381 Mdot
0.44283 0.17382 Mdot
0.46946 0.18614 Mdot
0.49226 0.19789 Mdot
0.51514 0.21025 Mdot
0.53943 0.22289 Mdot
0.55838 0.23255 Mdot
0.58256 0.24686 Mdot
0.60809 0.26171 Mdot
0.63254 0.27829 Mdot
0.65361 0.29154 Mdot
0.6793 0.30795 Mdot
0.70026 0.32368 Mdot
0.72606 0.34257 Mdot
0.74805 0.3606 Mdot
0.76999 0.37746 Mdot
0.79193 0.39891 Mdot
0.82024 0.42463 Mdot
0.84316 0.44671 Mdot
0.86607 0.46598 Mdot
0.88902 0.49375 Mdot
0.91284 0.51605 Mdot
0.93675 0.54443 Mdot
0.95515 0.56288 Mdot
grestore
grestore
grestore
MathPictureEnd


MathPictureStart
0.02381 0.12956 0.22236 170.114 [
[(Fig. 2)] 0.5 0.62428 0 -1 Msboxa
[(1)] 0.15337 0.20986 0 1 Msboxa
[(2)] 0.28293 0.20986 0 1 Msboxa
[(3)] 0.41248 0.20986 0 1 Msboxa
[(4)] 0.54204 0.20986 0 1 Msboxa
[(5)] 0.6716 0.20986 0 1 Msboxa
[(6)] 0.80116 0.20986 0 1 Msboxa
[(7)] 0.93072 0.20986 0 1 Msboxa
[(Aspect Ratio)] 1.00625 0.22236 -1 0 Msboxa
[(-0.001)] 0.01131 0.05225 1 0 Msboxa
[(-0.0005)] 0.01131 0.1373 1 0 Msboxa
[(0.0005)] 0.01131 0.30742 1 0 Msboxa
[(0.001)] 0.01131 0.39247 1 0 Msboxa
[(0.0015)] 0.01131 0.47753 1 0 Msboxa
[(0.002)] 0.01131 0.56259 1 0 Msboxa
[(Deviation)] 0.02381 0.62428 0 -1 Msboxa
[ -0.001 -0.001 0 0 ]
[ 1.001 0.61903 0 0 ]
] MathScale
1 setlinecap
1 setlinejoin
newpath
[ ] 0 setdash
0 setgray
0 setgray
[(Fig. 2)] 0.5 0.62428 0 -1 Mshowa
gsave
gsave
0.002 setlinewidth
0 0.22236 moveto
1 0.22236 lineto
stroke
0.15337 0.21611 moveto
0.15337 0.22861 lineto
stroke
0 setgray
[(1)] 0.15337 0.20986 0 1 Mshowa
0.28293 0.21611 moveto
0.28293 0.22861 lineto
stroke
0 setgray
[(2)] 0.28293 0.20986 0 1 Mshowa
0.41248 0.21611 moveto
0.41248 0.22861 lineto
stroke
0 setgray
[(3)] 0.41248 0.20986 0 1 Mshowa
0.54204 0.21611 moveto
0.54204 0.22861 lineto
stroke
0 setgray
[(4)] 0.54204 0.20986 0 1 Mshowa
0.6716 0.21611 moveto
0.6716 0.22861 lineto
stroke
0 setgray
[(5)] 0.6716 0.20986 0 1 Mshowa
0.80116 0.21611 moveto
0.80116 0.22861 lineto
stroke
0 setgray
[(6)] 0.80116 0.20986 0 1 Mshowa
0.93072 0.21611 moveto
0.93072 0.22861 lineto
stroke
0 setgray
[(7)] 0.93072 0.20986 0 1 Mshowa
0 setgray
[(Aspect Ratio)] 1.00625 0.22236 -1 0 Mshowa
0.02381 0 moveto
0.02381 0.61803 lineto
stroke
0.01756 0.05225 moveto
0.03006 0.05225 lineto
stroke
0 setgray
[(-0.001)] 0.01131 0.05225 1 0 Mshowa
0.01756 0.1373 moveto
0.03006 0.1373 lineto
stroke
0 setgray
[(-0.0005)] 0.01131 0.1373 1 0 Mshowa
0.01756 0.30742 moveto
0.03006 0.30742 lineto
stroke
0 setgray
[(0.0005)] 0.01131 0.30742 1 0 Mshowa
0.01756 0.39247 moveto
0.03006 0.39247 lineto
stroke
0 setgray
[(0.001)] 0.01131 0.39247 1 0 Mshowa
0.01756 0.47753 moveto
0.03006 0.47753 lineto
stroke
0 setgray
[(0.0015)] 0.01131 0.47753 1 0 Mshowa
0.01756 0.56259 moveto
0.03006 0.56259 lineto
stroke
0 setgray
[(0.002)] 0.01131 0.56259 1 0 Mshowa
0 setgray
[(Deviation)] 0.02381 0.62428 0 -1 Mshowa
grestore
grestore
0 0 moveto
1 0 lineto
1 0.618034 lineto
0 0.618034 lineto
closepath
clip
newpath
0 setgray
gsave
gsave
gsave
0.008 setlinewidth
0.15337 0.34484 Mdot
0.15997 0.36768 Mdot
0.16697 0.40815 Mdot
0.17423 0.01472 Mdot
0.182 0.38282 Mdot
0.19094 0.44927 Mdot
0.19858 0.56027 Mdot
0.20739 0.32299 Mdot
0.21659 0.48208 Mdot
0.22618 0.33117 Mdot
0.23719 0.45379 Mdot
0.24794 0.51158 Mdot
0.26012 0.39855 Mdot
0.27127 0.55304 Mdot
0.28474 0.54037 Mdot
0.29899 0.53407 Mdot
0.31195 0.56583 Mdot
0.32646 0.43897 Mdot
0.34162 0.54556 Mdot
0.36027 0.60332 Mdot
0.37711 0.50023 Mdot
0.39486 0.39377 Mdot
0.41469 0.42981 Mdot
0.43088 0.51473 Mdot
0.45252 0.44549 Mdot
0.47661 0.46134 Mdot
0.50097 0.35237 Mdot
0.523 0.37382 Mdot
0.55124 0.40946 Mdot
0.57547 0.35862 Mdot
0.60682 0.35108 Mdot
0.63493 0.31444 Mdot
0.66434 0.32292 Mdot
0.69518 0.26774 Mdot
0.73716 0.25838 Mdot
0.77304 0.25073 Mdot
0.81074 0.26744 Mdot
0.85039 0.24121 Mdot
0.89366 0.24939 Mdot
0.9394 0.24084 Mdot
0.97619 0.24325 Mdot
grestore
grestore
gsave
gsave
0.008 setlinewidth
0.15337 0.16112 Mdot
0.15997 0.12272 Mdot
0.16697 0.22613 Mdot
0.17423 0.34474 Mdot
0.182 0.21441 Mdot
0.19094 0.30127 Mdot
0.19858 0.19283 Mdot
0.20739 0.1937 Mdot
0.21659 0.30517 Mdot
0.22618 0.32267 Mdot
0.23719 0.33812 Mdot
0.24794 0.45205 Mdot
0.26012 0.29818 Mdot
0.27127 0.27575 Mdot
0.28474 0.40598 Mdot
0.29899 0.41499 Mdot
0.31195 0.35829 Mdot
0.32646 0.41685 Mdot
0.34162 0.3023 Mdot
0.36027 0.28521 Mdot
0.37711 0.34543 Mdot
0.39486 0.4397 Mdot
0.41469 0.43492 Mdot
0.43088 0.33271 Mdot
0.45252 0.29409 Mdot
0.47661 0.30994 Mdot
0.50097 0.26221 Mdot
0.523 0.30067 Mdot
0.55124 0.33291 Mdot
0.57547 0.30759 Mdot
0.60682 0.29835 Mdot
0.63493 0.28042 Mdot
0.66434 0.24807 Mdot
0.69518 0.25072 Mdot
0.73716 0.22776 Mdot
0.77304 0.24903 Mdot
0.81074 0.23682 Mdot
0.85039 0.23611 Mdot
0.89366 0.23238 Mdot
0.9394 0.23063 Mdot
0.97619 0.23134 Mdot
grestore
grestore
grestore
MathPictureEnd